\documentclass[twocolumn, prr, superscriptaddress]{revtex4-2}
\usepackage{amssymb}
\usepackage{amsmath}
\usepackage{epsfig}
\usepackage{color}
\usepackage{graphics, graphicx}
\usepackage{bbold}
\usepackage{psfrag}
\usepackage{mathcomp}
\usepackage{subfigure}
\usepackage{verbatim}
\usepackage{color}
\usepackage[colorlinks,citecolor=blue]{hyperref}

\begin{document}
\title{Angular topological superfluid and topological vortex in an ultracold Fermi gas}

\author{Ke-Ji Chen}
\thanks{These authors contributed equally to this work.}
\affiliation{Key Laboratory of Optical Field Manipulation of Zhejiang Province and Physics Department of Zhejiang Sci-Tech University, Hangzhou 310018, China}

\author{Fan Wu}
\thanks{These authors contributed equally to this work.}
\affiliation{Fujian Key Laboratory of Quantum Information and Quantum Optics, College of Physics and Information Engineering, Fuzhou University, Fuzhou, Fujian 350108, China}

\author{Lianyi He}
\email{lianyi@mail.tsinghua.edu.cn}
\affiliation{Department of Physics and State Key Laboratory of Low-Dimensional
Quantum Physics, Tsinghua University, Beijing 100084, China}

\author{Wei Yi}
\email{wyiz@ustc.edu.cn}
\affiliation{CAS Key Laboratory of Quantum Information, University of Science and Technology of China, Hefei 230026, China}
\affiliation{CAS Center For Excellence in Quantum Information and Quantum Physics, Hefei 230026, China}

\begin{abstract}
We show that pairing in an ultracold Fermi gas under spin-orbital-angular-momentum coupling (SOAMC) can acquire topological characters encoded in the quantized angular degrees of freedom. The resulting topological superfluid is the angular analog of its counterpart in a one-dimensional Fermi gas with spin-orbit coupling, but characterized by a Zak phase defined in the angular-momentum space. Upon tuning the SOAMC parameters, a topological phase transition occurs, which is accompanied by the closing of the quasiparticle excitation gap. Remarkably, a topological vortex state can also be stabilized by deforming the Fermi surface, which is topologically non-trivial in both the coordinate and angular-momentum space, offering interesting potentials for applications in quantum information and quantum control. We discuss how the topological phase transition and the exotic vortex state can be detected experimentally.
\end{abstract}

\maketitle

\section{introduction}
Synthetic gauge fields such as spin-orbit coupling (SOC) play a prominent role in quantum simulations using cold atoms~\cite{Lin-11, Zhang-12, Zwierlein-12, socreview1, socreview2, socreview3, socreview4, socreview5, socreview6}.
Under a synthetic SOC, for instance, different hyperfine ground states of the atoms are coupled optically, where a spin flip is accompanied by the change of the atom's center-of-mass momentum. The coupling thus modifies the atom's single-particle dispersion, which, while capable of inducing exotic few- and many-body quantum states~\cite{puhantwobody, shenoy,  soc3body1, soc3body2, Wu-13, chuanwei-13, Wei-13, tfflo0, tfflohu}, can also give rise to topological band structures~\cite{Qi-16, Shuai-16, Zhang-16,xjliu1D,xjliu2D}. The non-trivial band topology further lays the ground for topological superfluids---for instance, a topological superfluid emerges in a two-dimensional Fermi gas under the Rashba-type SOC, aided by Zeeman fields and pairing interactions~\cite{Kane-08,tsfsolid1,tsfsolid2,tsfsolid3,tsfsolid4,soc1,Yi-11}. Therein, the SOC and Zeeman fields mix different spin species and open up a gap between dressed single-particle bands, as a chiral $p+ip$ pairing order is induced out of an $s$-wave pairing interaction~\cite{Kane-08}. Notably, the topological superfluid is characterized by non-Abelian quasiparticle excitations that are potentially useful for quantum computation.

Recently, an alternative type of synthetic gauge field,  spin-orbital-angular-momentum coupling (SOAMC), has attracted considerable attention, where the atomic hyperfine spins are coupled to the center-of-mass angular momentum through a Raman process~\cite{Hu-15, Pu-15, Qu-15, Sun-15, Lin-18, Jiang-19}. As the Raman lasers are
Laguerre-Gaussian beams with different orbital angular momenta, this angular-momentum difference of light is imprinted onto the hyperfine spins of each single atom, with
profound many-body implications~\cite{Hu-15, Pu-15, Qu-15, Sun-15, Chen-16, Hu-19, Chen-19, Han-20, Duan-20}. For instance, while the SOAMC-driven vortex formation and phase transitions have been recently observed in Bose-Einstein condensates~\cite{Lin-18, Jiang-19}, theoretical studies reveal that the interplay of SOAMC and pairing interactions underlies a unique vortex-forming mechanism in Fermi superfluids~\cite{Chen-20, Wang-21}.
Here, a series of intriguing questions arise, whether topological superfluids can also be stabilized under SOAMC, and if so, in what form.

Naively, since the SOAMC is the angular analog of the conventional SOC, one would expect that a topological superfluid should be stabilized under similar conditions, with both pairing order and topology emerging in the angular-momentum space. The resulting angular topological superfluid should be understood on the same basis as the topological superfluid in a one-dimensional lattice gas under a one-dimensional SOC, with quantized angular momenta in the former playing the role of discretized linear momenta of the latter. However, implementing SOAMC relies on the spatial dependence of the Laguerre-Gaussian beams, such that the atomic gas must have spatial dimensions higher than one; whereas it is known that Fermi superfluids become gapless and lose their topological features when the spatial dimension becomes higher than that of the SOC~\cite{Wu-13,yisoc}. Given the one-dimensional nature of the SOAMC (coupling only occurs along the azimuthal direction), this imposes a stringent constraint on the stability of an angular topological superfluid.

In this paper, we show that a fully gapped angular topological superfluid survives the constraint above, provided the radial motion of the atoms is sufficiently suppressed such that it does not close the topological gap.
This can be experimentally achieved, for instance, by imposing strong Lagurre-Gaussian beams with a ring geometry, such that the atoms are tightly confined in the radial direction.
The topology of the system can then be captured by an effective one-dimensional topological model, where internal spin states are coupled to the quantized modes of the angular momentum. Under this configuration, we further demonstrate how exotic topological vortices can be engineered, whose topological features in both the coordinate and angular-momentum space offer intriguing potentials for robust quantum control.

\begin{figure}[tbp]
\begin{center}
\includegraphics[width=0.48\textwidth]{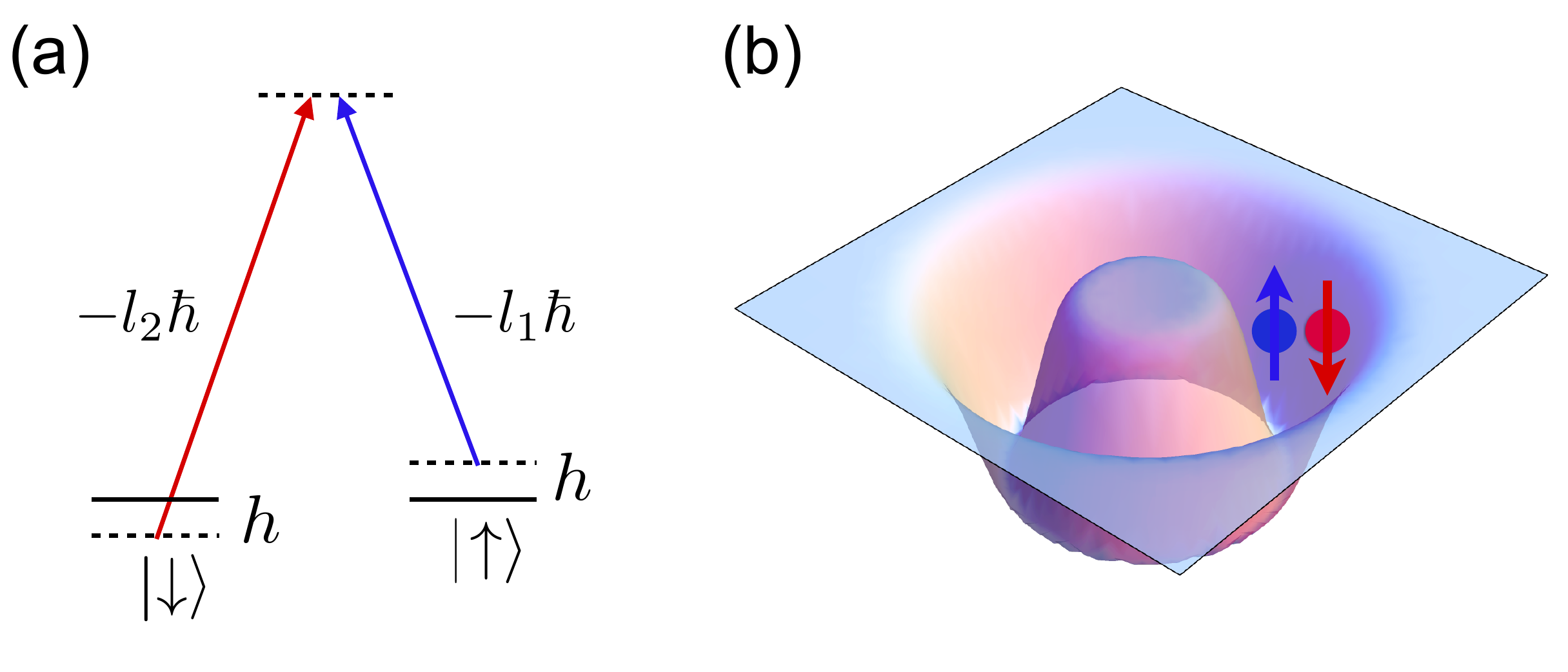}
\caption{(a) Schematic illustration of the level scheme. A pair of co-propagating Raman beams carrying different orbital angular momenta ($-l_1\hbar$ and $-l_2\hbar$) couple different hyperfine states, and transfer an angular momentum $2l\hbar=(l_1-l_2)\hbar$.
(b) Schematic illustration of atomic potentials under Lagurre-Gaussian beams with a ring geometry.}
\label{Fig1}
\end{center}
\end{figure}

\section{Model}
As illustrated in Fig.~\ref{Fig1}, we consider a two-component Fermi gas confined in the $x$--$y$ plane, where the two ground hyperfine states are labeled $\uparrow$ and $\downarrow$, respectively. The effective single-particle Hamiltonian can be written as
\begin{align}
\begin{split}
{\cal H}_s &=-\frac{\hbar^2}{2M}\frac{1}{r}\frac{\partial}{\partial r}\left(r\frac{\partial}{\partial r}\right)+\frac{\left(L_z-l \hbar \sigma_z\right)^2}{2Mr^2}  \\
& \ \ \ +\chi(r)+\Omega(r)\sigma_{x}+V_{\text{ext}}({\bf r}) -h \sigma_z,
\label{Hamiltonian-a}
\end{split}
\end{align}
where $M$ is the atom mass, $\sigma_i$ ($i=x,y,z$) are the Pauli matrices, $-l\hbar L_z\sigma_z/(Mr^2)$ is the SOAMC term, $L_{z}=-i\hbar \partial /\partial \theta$ is the atomic angular momentum operator, and $2l\hbar$ is the transferred orbital angular momentum from the co-propagating Raman beams. Here the polar coordinate ${\bf r}=(r,\theta)$ is taken. We also impose a hard-wall box potential $V_{\rm ext}({\bf r})$ with a radius $R$, to provide a natural boundary. Note that a gauge transformation $U=e^{-il\theta\sigma_z}$~\cite{Chen-20} is imposed to derive Hamiltonian (\ref{Hamiltonian-a}).

Consistent with previous experiments on SOAMC~\cite{Lin-18, Jiang-19}, the Raman coupling and the ac Stark potential are written as $\Omega(r)=\Omega_0 I(r)$ and $\chi(r)=\chi_0 I(r)$,
respectively. Here $\Omega_0$ is the effective two-photon Rabi frequency, $\chi_0$ is the ac Stark light shift, and $I(r)= (\sqrt{2}r/w)^{2l}e^{-2 r^2 /w^2}$ is the spatial intensity profile of the Laguerre-Gaussian lasers, with $w$ the beam waist. As we show below, for a confinement that is sufficiently tight along the radial direction, the radial degrees of freedom of the atoms can be frozen,
and the remaining quantized angular motion can be well-captured by an effective one-dimensional model with discretized modes. Intuitively, such a scenario occurs when the trap depth $\chi_0$ is large, so that the radial excitation energy ($\sim \sqrt{ |\chi_0| \hbar^2 /(m w^2)}$, see Appendix \ref{app:radial}) becomes much larger than any other relevant energy scales of the system. The atoms are then localized near $r_0=\sqrt{l/2}w$ in the radial direction. Under typical experimental conditions (taking $^{6}$Li atoms as an example), with the laser waist $w\sim 7.5$ $ \mu m$ and a relatively large $l$, we estimate the ac Stark shift $|\chi_0|/(2\pi \hbar)$ to be on the order of hundreds of Hz, which is experimentally achievable.

For the pairing physics, we consider an $s$-wave contact interaction between different spin species. The full Hamiltonian is written as
$H=H_0+H_{\text{int}}$, with $H_0=\int d{\bf r} \Psi^{\dag}({\bf r}) {\cal{H}}_s \Psi({\bf r})$ and $H_{\text{int}}=-g\int d{\bf r}\psi^{\dag}_{\uparrow}({\bf r})\psi^{\dag}_{\downarrow}({\bf r})\psi_{\downarrow}({\bf r})\psi_{\uparrow}({\bf r})$. We write $\Psi(\bf r)=[\psi_\uparrow(\bf r),\psi_\downarrow(\bf r)]^{\rm T}$, where $\psi_{\sigma}(\bf r)$ ($\sigma=\uparrow,\downarrow$) are the field operators for the two spin species. The bare interaction rate $g$ is renormalized by relating it to the two-body bound-state energy $E_B$ in the same ac Stark potential but without the spin-mixing Raman term $\Omega(r)$   (see Appendix \ref{app:renormal}).

We study the many-body pairing physics using the Bogoliubov--de Gennes (BdG) formalism (see Appendix \ref{app:BdG} ), where the BdG equation is given by ${\cal H}_{\text{BdG}}\Phi_{mn}({\bf r})=\epsilon_{mn}\Phi_{mn}({\bf r})$, with $\epsilon_{mn}$ the Bogoliubov spectrum, and $m$ and $n$ the angular and radial quantum number, respectively. Here
\begin{align}
\cal H_{\text{BdG}}
=\left[\begin{array}{cccc}K_{\uparrow }({\bf r}) &\Omega (r) & 0 & \Delta({\bf r}) \\\Omega (r) & K_{\downarrow }({\bf r}) & -\Delta({\bf r}) & 0 \\0 & -\Delta^{\ast}({\bf r}) & -K^{\ast}_{\uparrow}({\bf r}) & -\Omega (r) \\ \Delta^{\ast}({\bf r}) & 0 & -\Omega (r) & -K^{\ast}_{\downarrow}({\bf r})\end{array}\right],
\label{HC}
\end{align}
and $\Phi_{mn}({\bf r})=[ u_{\uparrow mn}, u_{\downarrow mn} , v_{\uparrow mn} , v_{\downarrow mn}]^{\rm T}$. The Bogoliubov coefficients $u_{\sigma mn}$ and $v_{\sigma mn}$ are now spatially dependent under the ring geometry of the system. The pairing order parameter $\Delta({\bf r})=-g \langle \psi_{\downarrow}({\bf r}) \psi_{\uparrow}({\bf r})\rangle$ is expressed as $\Delta({\bf r})   =  \frac{g}{2}\sum_{mn} [u_{\uparrow mn}v^* _{\downarrow mn}  \vartheta (\epsilon_{mn})+u_{\downarrow mn} v^*_{\uparrow mn }\vartheta(-\epsilon_{mn}) ]$ [$\vartheta (x)$ is the Heaviside step function]. We also have $K_{\sigma}({\bf r})  ={\cal K}_{\sigma}-\mu_{\sigma}$, with $\mu_{\sigma}=\mu+h\tau$ and $\mu$ being the chemical potential, and ${\cal K}_{\sigma}= -\frac{\hbar^2}{2M} \left[\frac{1}{r}\frac{\partial}{\partial r}(r\frac{\partial}{\partial r})+\frac{1}{r^2}\left(\frac{\partial}{\partial \theta}-i\tau l \right)^2 \right]+\chi(r)$, with $\tau=+1\ (-1)$ for $\sigma=\uparrow(\downarrow)$.

To solve the many-body ground state, we write the order parameter as $\Delta({\bf r})=\Delta(r)e^{i \kappa \theta}$, where the vorticity $\kappa=0$ ($\kappa \neq 0$) indicates a vortexless superfluid state (vortex  state). Starting with different values of $\kappa \in \mathbb{Z}$, we self-consistently solve the BdG equation under the constraint $N=\sum_\sigma\int d{\bf  r} n_\sigma({\bf r})$, where $N$ is the total particle number, and the density distribution $n_{\sigma}({\bf r})  =  \frac{1}{2}\sum_{mn}[ |u_{\sigma mn}|^2\vartheta(-\epsilon_{mn})+|v_{\sigma mn}|^2 \vartheta(\epsilon_{mn})]$. The ground state of the system is determined by comparing free energies $F$ for different $\kappa$ (see Appendix \ref{app:free}).

\section{Angular topological superfluid}
\begin{figure}[tbp]
\begin{center}
\includegraphics[width=0.48\textwidth]{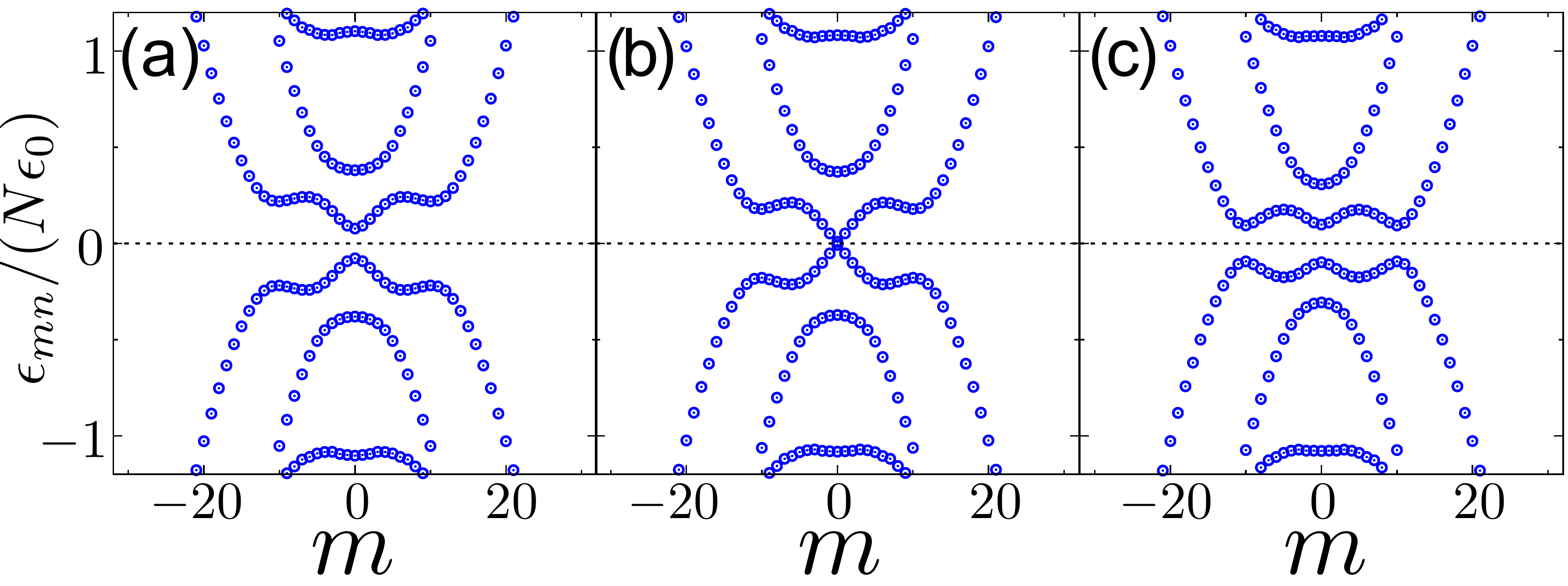}
\caption{Bogoliubov spectra of the Fermi superfluid under SOAMC, with a vanishing two-photon detuning $h=0$, and an increasing $\Omega_0$: (a) $\Omega_0/\epsilon_0=0.15$, (b) $\Omega_0/\epsilon_0=0.18$, and (c) $\Omega_0/\epsilon_0=0.2$. Other parameters are: $\kappa=0$, $l=5$, $R/w=5$,  $\chi_0/\epsilon_0=-8$, $E_B=2 E_{\rm min}-6\epsilon_0$, where $E_{\rm min}$ is the ground-state energy of ${\cal K}_{\sigma}$. }
\label{Fig2}
\end{center}
\end{figure}

We start by solving the Bogoliubov quasiparticle spectrum in a sufficiently deep ac Stark potential, with $\chi_0/\epsilon_0=-8$,  where we take $\epsilon_0=\pi^2 \hbar^2/(2M r^2_0)$ as the unit of energy.
For the case with $h=0$, we find that the ground state always lies with $\kappa=0$.
As illustrated in Fig.~\ref{Fig2}, by tuning the SOAMC parameter $\Omega_0$, the Bogoliubov spectrum undergoes a gap closing and re-opening process, reminiscent of that of a topological phase transition. Specifically, the Bogoliubov quasiparticle excitation is fully gapped under small $\Omega_0$ [Fig.~\ref{Fig2}(a)], becomes gapless at a critical $\Omega^{c}_0/\epsilon_0\approx 0.18 $ [Fig.~\ref{Fig2}(b)], and is again fully gapped for $\Omega_0>\Omega_0^c$ [Fig.~\ref{Fig2}(c)].

To provide a transparent picture of the topological nature of the gap-closing transition, we adopt a single-mode approximation $\psi_{\sigma}({\bf r})  \approx  \sum_{m}\phi_{m  \sigma}(r)\Theta_{m}(\theta)a_{m\sigma}$, where $\Theta_m(\theta)=e^{i m \theta}/\sqrt{2\pi}$, and $\phi_{m  \sigma}(r)$ is the normalized radial ground-state wave function satisfying
${\cal K}_{\sigma} ({\bf r}) \phi_{m \sigma}(r) \Theta_{m}(\theta) =  E_{m  \sigma}\phi_{m \sigma}(r) \Theta_{m}(\theta)$, with $E_{m  \sigma}$ the single-particle eigenenergies.
As we confirm below, the single-mode approximation works well, provided the atomic radial motion is frozen.

The effective one-dimensional Hamiltonian under the single-mode approximation becomes
\begin{align}
H_{\text{MF}} = &  \sum_{m \sigma}\xi_{m \sigma}a^{\dag}_{m \sigma}a_{m \sigma} +\sum_{m}(\Omega_{m}a^{\dag}_{m \uparrow}a_{m \downarrow}+H.c.) \nonumber \\
& +\sum_{m} (\Lambda^{\kappa}_m a^{\dag}_{m\uparrow}a^{\dag}_{-m+\kappa \downarrow}+H.c.),
\label{MF-H}
\end{align}
where $a_{m\sigma}$ ($a^\dag_{m\sigma}$) is the fermion annihilation (creation) operator for the corresponding spin species in the angular mode $m$, $\xi_{m \sigma}=E_{m \sigma}-\mu_{\sigma}$, and $\Omega_m =  \Omega_0 \int rdr \phi_{m \uparrow}I(r) \phi_{m  \downarrow}$. The pairing order parameter $\Lambda^{\kappa}_{m} =  \sum_{m'}U^{\kappa}_{m,m'}\langle a_{-m'+\kappa \downarrow}a_{m' \uparrow}\rangle$, where $U^{\kappa}_{m,m'}=-\frac{g}{2 \pi} \int rdr \phi_{m \uparrow}\phi_{-m+\kappa, \downarrow} \phi_{-m'+\kappa,\downarrow}\phi_{m' \uparrow}$, and $\kappa=0$ ($\kappa \neq 0$) denotes the superfluid (vortex) state.
The bare interaction rate $g$ should be renormalized under the single-mode approximation (see Appendix \ref{app:renormal}).

The effective Hamiltonian Eq.~(\ref{MF-H}) allows for a much simplified solution of the many-body ground state. Formally, the BdG Hamiltonian is now
\begin{align}
H_{m}=\left [\begin{array}{cccc}\xi_{m \uparrow} & \Omega_m & 0 & \Lambda^{\kappa}_{m} \\ \Omega_m &  \xi_{m \downarrow}& -\Lambda^{\kappa}_{-m+\kappa} & 0 \\0 & -\Lambda^{\kappa}_{-m+\kappa}& -\xi_{-m+\kappa \uparrow} & -\Omega_{-m+\kappa} \\ \Lambda^{\kappa}_m& 0 & -\Omega_{-m+\kappa} & -\xi_{-m+\kappa \downarrow}\end{array}\right],
\label{eq:Hm}
\end{align}
and one may proceed to solve the BdG equation $H_{m} \Phi_{ms}  =  \epsilon_{ms} \Phi_{ms}$, by
writing $\Phi_{ms}=[u^{\uparrow}_{ms}, u^{\downarrow}_{ms}, v^{\uparrow}_{ms}, v^{\downarrow}_{ms}]^{\rm T}$ and following the same self-consistent approach as before.  Here, $u^{\sigma}_{ms}$ and $v^{\sigma}_{ms}$ are the Bogoliubov coefficients under the single-mode approximation.

\begin{figure}[t]
\begin{center}
\includegraphics[width=0.48\textwidth]{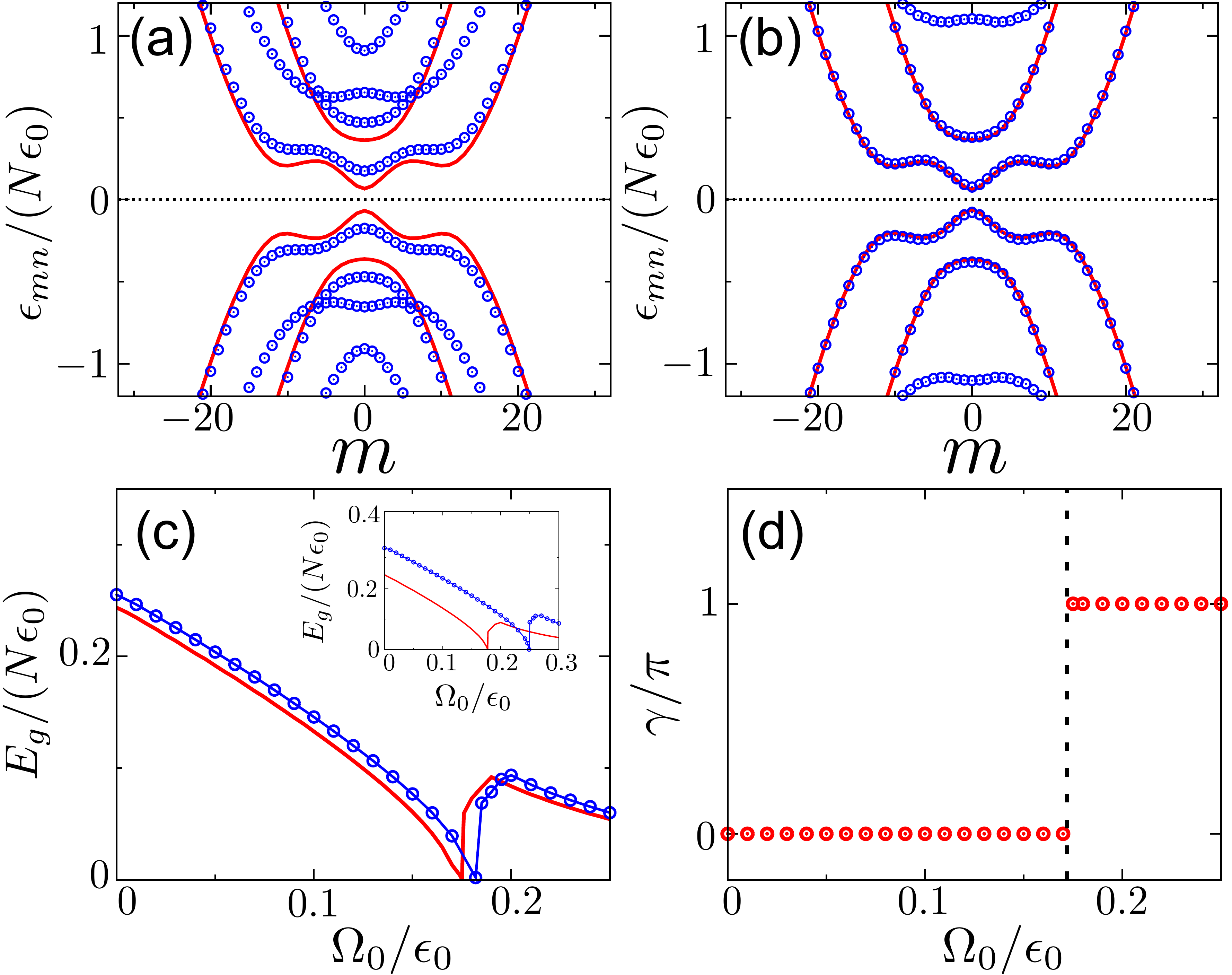}
\caption{Bogoliubov spectra from the full BdG calculation (blue symbols) and the single-mode approximation (red lines), with $\Omega_0/\epsilon_0=0.15$, for (a) $\chi_0/\epsilon_0=-3$ and (b) $\chi_0/\epsilon_0=-8$, respectively. (c) Variation of the quasiparticle excitation gap
with increasing $\Omega_0$, where the blue symbols and the red line respectively denote results from the full BdG calculation and the single-mode approximation. We take $\chi_0/\epsilon_0=-8$ here, while we take $\chi_0/\epsilon_0=-3$ for the inset.
(d) Topological transition revealed through the Zak phase of the effective Hamiltonian (\ref{eq:Hm}). The black dashed line indicate the gap-closing point. Other parameters are the same as those in Fig.~\ref{Fig2}.}
\label{Fig3}
\end{center}
\end{figure}

To demonstrate the validity of the single-mode approximation, we compare the Bogoliubov spectrum between the full BdG results from Hamiltonian (\ref{HC}) with those from (\ref{eq:Hm}). As shown in Fig.~\ref{Fig3}(a), when the ac Stark potential is only moderately deep (with $\chi_0/\epsilon_0=-3$),
there still appears to be an appreciable difference between the low-lying quasiparticle spectra. The difference is essentially gone for $\chi_0/\epsilon_0=-8$ [see Fig.~\ref{Fig3}(b)]. We further show the quasiparticle excitation gap in Fig.~\ref{Fig3}(c), where only a small shift in the gap-closing point is observed for $\chi_0/\epsilon_0=-8$, while a more significant shift is observed under $\chi_0/\epsilon_0=-3$ [inset of Fig.~\ref{Fig3}(c)]. Hence, while the single-mode approximation provides a satisfactory description for a deep ac Stark potential, topological phase transitions persist beyond the approximation.

The mean-field Hamiltonian Eq.~(\ref{MF-H}) is equivalent to the Bardeen-Cooper-Schrieffer Hamiltonian for a one-dimensional lattice gas under SOC, with the quantized angular momentum modes corresponding to discrete lattice momentum of the latter.  The topological invariant is given by
the Zak phase~\cite{Zak-89,Wei-12}
\begin{align}
\gamma= i \sum_{m} \Phi^{*}_{ms}   D_m \Phi_{ms},
\label{eq:zak}
\end{align}
with $D_{m} \Phi_{ms}=\Phi_{m+1,s}-\Phi_{ms}$ for $m$. The summation in Eq.~(\ref{eq:zak}) runs over the occupied states with $\epsilon_{ms}<0$.

In Fig.~\ref{Fig3}(d), we plot the Zak-phase variation with increasing $\Omega_0$. A topological phase transition is identified at the critical $\Omega^c_0$ where $\gamma$ changes from zero (topologically trivial) to $\pi$ (topologically non-trivial), where $\Omega^c_0$ agrees with the gap-closing point in Fig.~\ref{Fig2} under the same parameters.
The system is therefore in an angular topological superfluid for $\Omega_0>\Omega^c_0$,
reminiscent of the topological superfluid under SOC and $s$-wave pairing interaction~\cite{Liu-12, Wei-12}.

\begin{figure}[t]
\begin{center}
\includegraphics[width=0.48\textwidth]{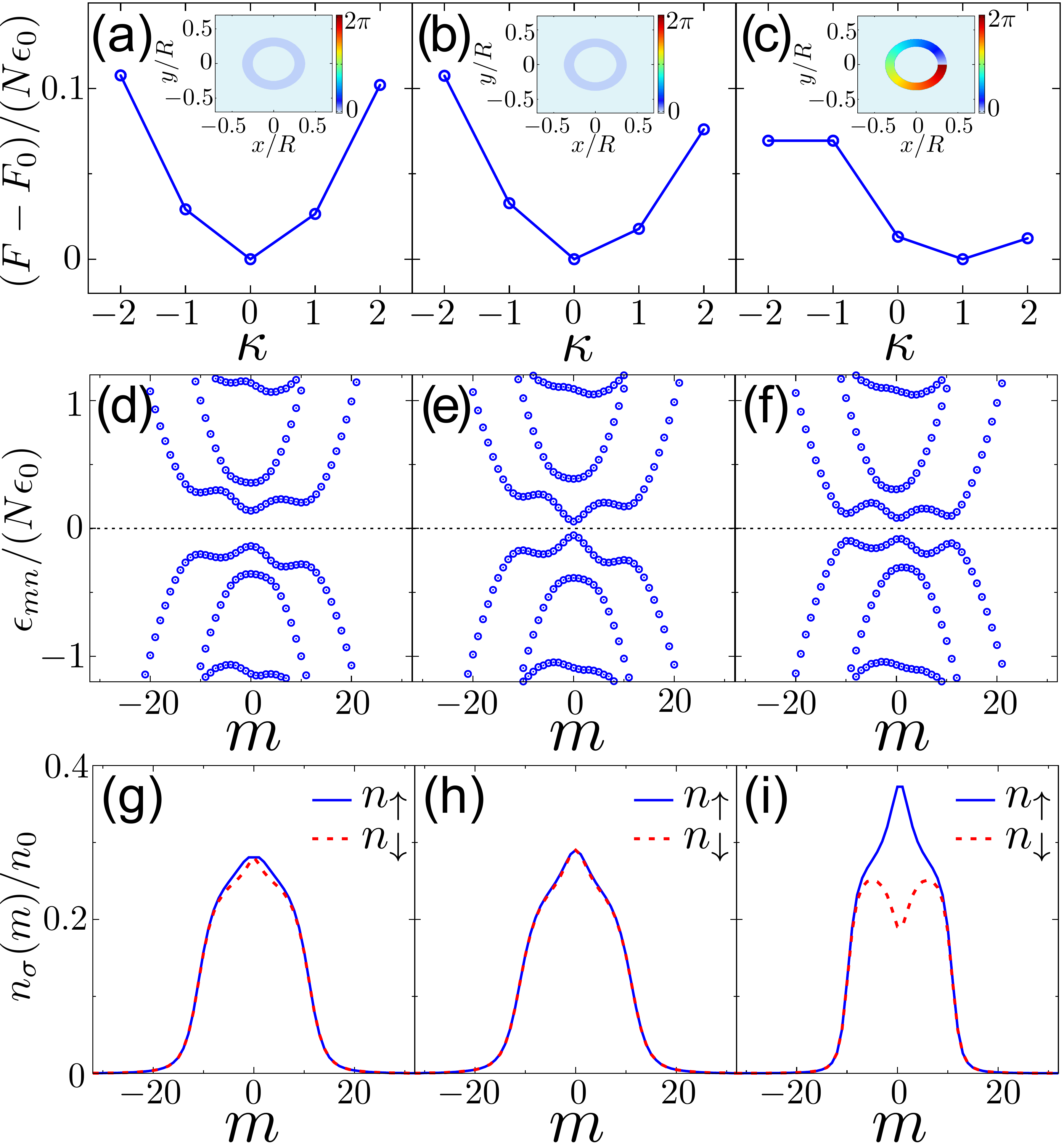}
\caption{(a)--(c) Free energies of pairing states as functions of $\kappa$, with a fixed two-photon detuning $h/\epsilon_0=0.8$. The insets show the phase of the order parameter. Here, $F_0$ denotes the ground-state free energy. (d)--(f) Bogloliubov spectra of the ground pairing state with increasing $\Omega_0$. (g)--(i) Density distribution of the ground state in the angular-momentum space, where we define $n_0=N/(\pi r^2_0)$. Here, the blue solid (red dashed) curve denotes the density distribution of spin-up (spin-down) component. (a), (d), (g) and (b), (e), (h) are the vortexless superfluid state with $\Omega_0/\epsilon_0=0.1$ and $0.16$, respectively. (c), (f), (i) is a topological vortex state with $\Omega_0/\epsilon_0=0.18$.
Other parameters are the same as those in Fig.~\ref{Fig2}. All results here are from the full BdG calculations.}
\label{Fig4}
\end{center}
\end{figure}

\section{Topological vortex state}
Building upon the topological superfluid state above, we now show that an exotic topological vortex state can be induced by turning on the two-photon detuning $h$ which deforms the Fermi surface.

As illustrated in Figs.~\ref{Fig4}(a)--\ref{Fig4}(c), under a finite $h$, the free energy is generically asymmetric with respect to $\kappa=0$. The asymmetry becomes more apparent with increasing $\Omega_0$, until the ground-state order parameter eventually acquires a finite phase with $\kappa\neq 0$.
Intriguingly, the transition into the vortex state is topological. As demonstrated in Figs.~\ref{Fig4}(d)--\ref{Fig4}(f), while the Fermi-surface deformation is reflected as the asymmetric spectral shape with respect to $m=0$, the closing and re-opening of the energy gap persist. Indeed, after the gap is reopened, the angular momentum of the ground state changes from $\kappa=0$ to $\kappa=1$, and the ground state simultaneously becomes topological, which is confirmed by the Zak-phase calculation.
Conceptually, such an exotic topological vortex state is the angular version of the topological Fulde-Ferrell state under the conventional SOC~\cite{chuanwei-13, Wei-13, tfflo0, tfflohu}.

The topological vortex leaves a direct signature in the angular-momentum-space density profile, which are illustrated in Figs.~\ref{Fig4}(g)--\ref{Fig4}(i).
In the only topological vortex state of Fig.~\ref{Fig4}(i), the density profile of the minority spin species exhibits a dip close to $\kappa/2$. Similar signatures have been identified in the topological Fulde-Ferrell state under SOC. Note that the spin polarization in the vortex state is a direct result of the two-photon detuning, which plays the role of an effective Zeeman field.

\begin{figure}[t]
\begin{center}
\includegraphics[width=0.5\textwidth]{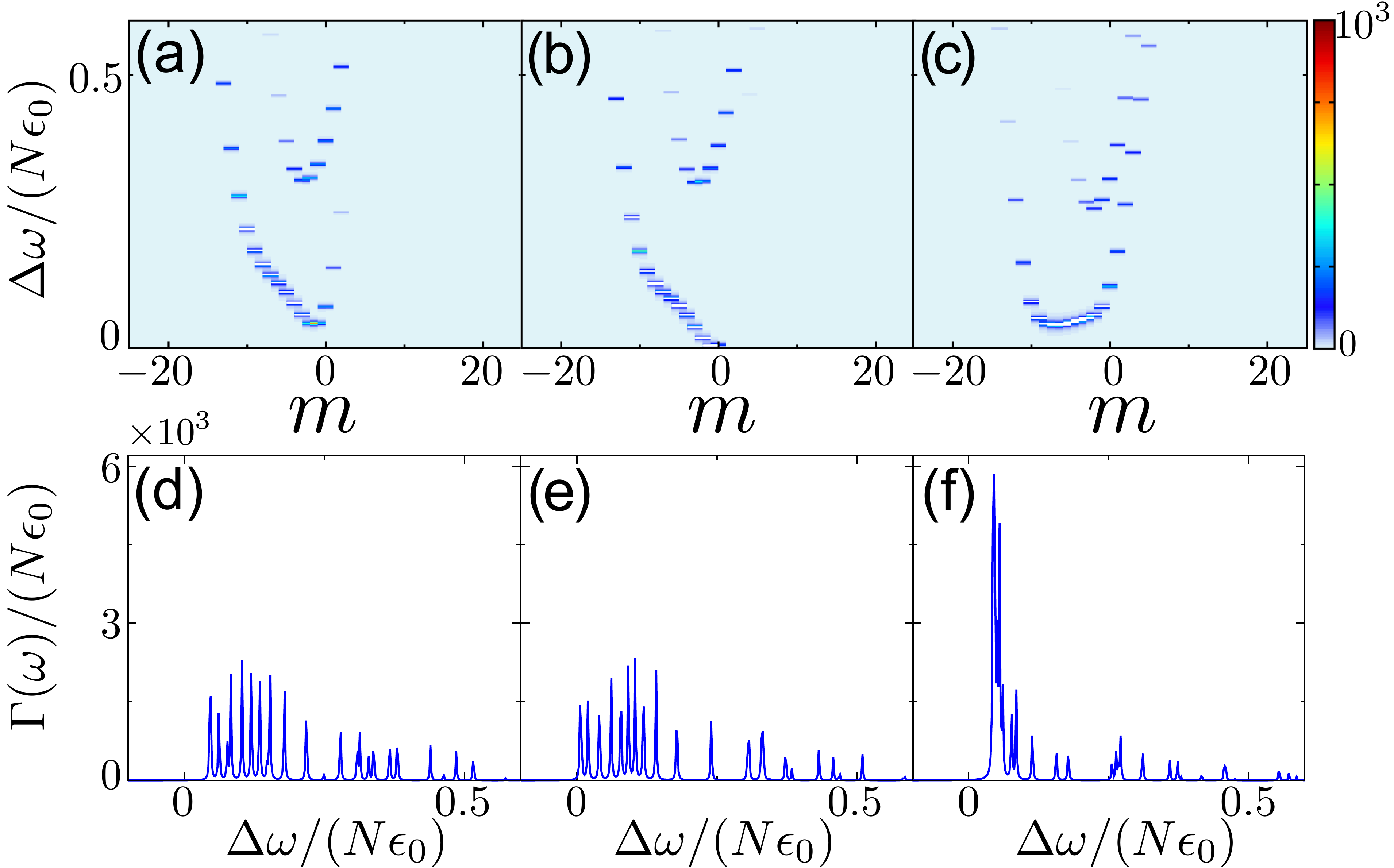}
\caption{(a)--(c) Contour  plots of the angular-momentum-resolved radio-frequency spectroscopy $A(m,\omega)$, when $\Omega_0$ increases under zero two-photon detuning. (d)--(f) The corresponding transition rate $\Gamma(\omega)$. Parameters for (a), (d), and (b), (e), and (c), (f) are $\Omega/\epsilon_0=0.15$, $0.18$, $0.21$, respectively. Here,  $\Delta \omega=\omega-E_l$. Other parameters are the same as those in Fig.~\ref{Fig2}.}
\label{Fig5}
\end{center}
\end{figure}

\section{Detecting topological transition}
The topological transition into the angular topological superfluid state can be detected through radio-frequency (rf) spectroscopy, by coupling one of the hyperfine states in the pairing superfluid (say $|\uparrow\rangle$) to a spectator state $|3\rangle$. The coupling Hamiltonian is
$V(t)=V_0 \int d{\bf r}e^{-i \omega t}\psi^{\dag}_{3}(\theta)e^{-i l \theta} \psi_{\uparrow}({\bf r})$, where $V_0$ is the coupling strength, $\psi^\dag_3$ is the creation operator of state $|3\rangle$, $\omega$ is the detuning of the rf pulse relative to the $|\uparrow\rangle\rightarrow |3\rangle$ transition, and the phase $e^{-i l \theta}$ comes from the gauge transformation under Eq.~(\ref{Hamiltonian-a}).
From the Fermi's golden rule, the transition rate to the spectator state is $\Gamma(\omega)=\sum_{m}A(m,\omega )$,
where
\begin{align}
A(m,\omega)=
\frac{2\pi V^2_0 }{\hbar} \Big[\sum_{n} \int d r v^{\uparrow}_{mn}(r) \Big]^2 \delta(\hbar \omega-\epsilon_{mn}-E_{m+l}).
\label{Gamma-m}
\end{align}
Here,  $v^{\uparrow}_{mn}(r) $ is the radial part of $v_{\uparrow mn}({\bf r})$,  with $v_{\uparrow mn}({\bf r})=v^{\uparrow}_{mn}(r)\Theta_{m-\kappa}(\theta)$,  and $E_{m}=m^2\epsilon_0/\pi^2$. We assume that atoms in state $|3\rangle$
remain trapped close to $r_0$, which is a good approximation given a sufficiently strong potential.
In Fig.~\ref{Fig5}, we show both the angular-momentum-resolved rf spectrum $A(m,\omega)$ and the full spectrum $\Gamma(\omega)$. As the SOAMC strength $\Omega_0$ is tuned, the gap-closing and reopening process is clearly identified, with Figs.~\ref{Fig5}(b)and \ref{Fig5}(e) corresponding to the gap-closing point.

\section{Discussion and outlook}
We show that an exotic angular topological superfluid can be stabilized in a Fermi superfluid under SOAMC, provided the atomic radial motion is suppressed. As topological superfluids are known to host Majorana zero modes at boundaries, similar zero modes should be observed in an angular topological superfluid, once a boundary is created, for instance, by shining a strong laser beam to break the ring geometry of the ac Stark potential. It would be even more interesting to generate such non-Abelian quasiparticle excitations in a topological vortex state whose simultaneous topology in the coordinate and angular-momentum space would make the zero modes doubly robust, and easier to control, for instance, through lasers generating the SOAMC. Our work thus offers an interesting alternative for the generation and control of Majorana zero modes in cold atoms.

\section{acknowledgments}
We acknowledge fruitful discussions with Jian-Song Pan, Li-Jun Lang, Dongyang Yu, Long Zhang, and Liang-Liang Wang. This work is supported by  the Natural Science Foundation of China (Grants No. 12104406, and No. 11974331) and the National Key R\&D Program (Grants No. 2018YFA0306503, and No. 2017YFA0304100). K.C. acknowledges support from the startup grant of Zhejiang Sci-Tech University (Grant No. 21062338-Y).  F. W. also acknowledges support from the startup grant of Fuzhou University  (Grant No. GXRC21065).

\section*{APPENDIX}
In these appendixes, we provide details on the radial excitation energy, renormalization of the bare interaction strength, the BdG formalism, and the expression of the free energy.

\appendix
\section{Radial excitation energy}
\label{app:radial}

To estimate the radial excitation energy, we expand the ac Stark potential $\chi(r)$ around its minimum at $r_0$ to the quadratic order
\begin{align}
\chi(r) \approx \chi_0  \Big[e^{-l}l^{l}-\frac{4}{w^2}(e^{-l}l^l)(r-r_0)^2 \Big].
\end{align}
The harmonic trapping frequency can then be written as $ \omega_{ho}=(8e^{-l}l^{l}  |\chi_0|/(m w^2))^{1/2}$, so that
the radial excitation energy is given by
\begin{align}
\hbar \omega_{ho}=\sqrt{8e^{-l}l^{l} \frac{|\chi_0| \hbar^2 }{m w^2}} \propto \sqrt{\frac{|\chi_0| \hbar^2}{m w^2 }}.
\end{align}

\section{Renormalization of Full Hamiltonian}
\label{app:renormal}

The renormalization relation can be obtained by solving a two-body problem in the absence of SOAMC, but within the same ac Stark potential. The corresponding Hamiltonian reads
\begin{align}
\begin{split}
{\cal H}  = &\sum_{\sigma} \int d {\bf r} \psi^{\dag}_{\sigma}({\bf r}) {\cal K}_{\sigma}({\bf r}) \psi_{\sigma}({\bf r})\\
&-g\int  d{\bf r} \psi^{\dag}_{\uparrow}({\bf r})\psi^{\dag}_{\downarrow}({\bf r})\psi_{\downarrow}({\bf r})\psi_{\uparrow}({\bf r}),
\end{split}
\end{align}
where ${\cal K}_{\sigma}({\bf r})$ is shown in the main text.

We expand the field operator $\psi_{\sigma}({\bf r})$ as $\psi_{\sigma}({\bf r})=\sum_{nm}\phi_{nm \sigma}(r)\Theta_{m}(\theta)a_{nm \sigma}$. Here, $\phi_{nm\sigma}(r) \Theta_{m}(\theta)$ is the eigen wavefunction of the operator ${\cal K}_{\sigma}({\bf r})$, with
$ {\cal K}_{\sigma}({\bf r})\phi_{nm \sigma}(r)\Theta_{m}(\theta)  = E_{nm \sigma} \phi_{nm \sigma}(r) \Theta_{m}(\theta)$, and $E_{nm\sigma}$ the eigenvalue.
The Hamiltonian is then
\begin{widetext}
\begin{eqnarray}
{\cal H}  =  \sum_{nm\sigma} E_{nm \sigma}a^{\dag}_{nm \sigma}a_{nm \sigma}  -\frac{g}{2\pi}\sum_{nn'n''n'''}\sum_{mm'm''} f^{n,n',n'',n'''}_{m,m',m''} a^{\dag}_{n''',m+m'-m''\uparrow}a^{\dag}_{n''m''\downarrow}a_{n'm' \downarrow}a_{n m \uparrow},
\end{eqnarray}
\end{widetext}
with $f^{n,n',n'',n'''}_{m,m',m''}=\int r dr \phi_{n''',m+m'-m'' \uparrow} \phi_{n''m'' \downarrow}\phi_{n'm'\downarrow}\phi_{nm \uparrow}$ the overlap of radial wavefunctions.

To solve the two-body problem, we write the two-body bound state as $|\Phi_{2B}
\rangle=\sum_{nm}\Phi_{nm} a^{\dag}_{nm \uparrow}a^{\dag}_{n,-m, \downarrow}| {\rm vac}\rangle$.
In doing so, we have neglected pairing between different radial modes, indexed by $n$. This should be a fair approximation, as different radial modes have distinct symmetries and hence a smaller overlap compared to that between the same modes. Hence pairing between different radial modes is suppressed due to the limited phase space.
From the Schr\"odinger equation ${\cal H} |\Phi_{2B} \rangle= E_{B} |\Phi_{2B} \rangle$,  we have
\begin{align}
&\sum_{nm }(2E_{nm \uparrow}-E_{B})\Phi_{nm }a^{\dag}_{nm \uparrow}a^{\dag}_{n,-m \downarrow}| {\rm vac} \rangle \nonumber\\
 &  =  \frac{g}{2\pi}\sum_{nm}\beta_{nm}a^{\dag}_{nm \uparrow}a^{\dag}_{n,-m \downarrow}| {\rm vac}\rangle,
\end{align}
where $E_{B}$ is the bound-state energy.
It follows that
\begin{eqnarray}
\Phi_{nm} & = & \frac{g}{2\pi} \frac{\beta_{nm}}{2E_{nm \uparrow}-E_{B}},
\label{Psi-nm}
\end{eqnarray}
where
\begin{eqnarray}
\beta_{nm} &=&\sum_{n'm'}\int rdr |\phi_{nm \uparrow}|^2 |\phi_{n'm' \uparrow}|^2\Phi_{n'm'}.\label{eq:beta}
\end{eqnarray}
Equation ~(\ref{Psi-nm}), together with Eq.~(\ref{eq:beta}), constitute a set of linear equations for $\Phi_{nm}$. A relation between $E_B$ and $g$ can be numerically established, by sending the determinant of the coefficient matrix to zero. This is the renormalization condition for the full BdG calculation.

Under the single-mode approximation, $\beta_{nm}$, $\Phi_{nm}$ and $E_{nm\sigma}$ reduce to $\beta_m$, $\Phi_m$ and $E_{m \sigma}$ (as defined in the main text), respectively. We then have
\begin{align}
\Phi_m & = \frac{g}{2\pi} \frac{\beta_{m}}{2E_{m \uparrow}-E_B},\label{beta1} 
\end{align}
\begin{align}
\beta_m & =  \sum_{m'} \int rdr |\phi_{m' \uparrow}|^2 |\phi_{m \uparrow}|^2 \Phi_{m'}.
\label{beta2}
\end{align}
A relation between $E_B$ and $g$ can be similarly established by sending the determinant of the coefficient matrix of Eq.~(\ref{beta1}) to zero. This is the renormalization relation under the single-mode approximation. Note that, in the limit of a very deep ac Stark potential, $\beta_m$ in Eq.~(\ref{beta2}) becomes independent of $m$, and a closed form of the renormalization relation can be obtained.

\section{Details of the BdG formalism}
\label{app:BdG}
As discussed in the main text, we assume $\Delta({\bf r})=\Delta(r)e^{i \kappa \theta}$ with $\kappa$ an integer. The Bogoliubov coefficients $u_{\sigma m n}$ and $v_{\sigma m n}$ can be expanded as
 \begin{eqnarray}
  u_{\sigma m n} &=& \sum_{n'}c^{(n')}_{\sigma mn}R_{n', m-l \tau}(r)\Theta_{m}(\theta), \label{u_sigma}\\
  v_{\sigma m n} &= &\sum_{n'}d^{(n')}_{\sigma mn}R_{n',m+l\tau-\kappa}(r)\Theta_{m-\kappa}(\theta),
\label{v_sigma}
\end{eqnarray}
where $R_{nm}(r)$ is the radial wave function of a two-dimensional system with a hard-wall potential, with $R_{nm}(r) = \sqrt{2}J_{m}(\alpha_{nm}\frac{r}{R})/R J_{m+1}(\alpha_{nm})$. Here,  $J_m(x)$ is the Bessel function of the first kind, whose zeros are given by $\alpha_{nm}$.

Substituting Eqs.~(\ref{u_sigma}) and ~(\ref{v_sigma}) into the BdG equation, we determine $c^{(n')}_{\sigma mn}$ and $d^{(n')}_{\sigma mn}$. For each $m$, we have
\begin{widetext}
\begin{eqnarray}
  \sum_{n''}
  \left[
    \begin{array}{cccc}
      K^{n'n''}_{\uparrow,m-l} & \Omega^{n'',m+l}_{n',m-l} & 0 & \Delta^{n'',m-l-\kappa}_{n',m-l} \\
      \Omega^{n'',m-l}_{n',m+l}& K^{n'n''}_{\downarrow,m+l} & -\Delta^{n'',m+l-\kappa}_{n',m+l}& 0 \\
      0 & -\Delta^{n'',m+l}_{n',m+l-\kappa} & -K^{n'n''}_{\uparrow,m+l-\kappa} & -\Omega^{n'',m-l-\kappa}_{n',m+l-\kappa} \\
      \Delta^{n'',m-l}_{n',m-l-\kappa} & 0 & -\Omega^{n'',m+l-\kappa}_{n',m-l-\kappa} & -K^{n'n''}_{\downarrow, m-l-\kappa} \\
    \end{array}
  \right]
  \left[
    \begin{array}{c}
      c^{(n'')}_{\uparrow mn} \\
      c^{(n'')}_{\downarrow mn} \\
      d^{(n'')}_{\uparrow mn} \\
      d^{(n'')}_{\downarrow mn} \\
    \end{array}
  \right]
  =\epsilon_{mn}
  \left[
    \begin{array}{c}
      c^{(n')}_{\uparrow mn} \\
      c^{(n')}_{\downarrow mn} \\
      d^{(n')}_{\uparrow mn} \\
      d^{(n')}_{\downarrow mn} \\
    \end{array}
  \right],
\label{BDG1}
\end{eqnarray}
where
\begin{eqnarray}
K^{n'n''}_{\sigma, p}  &=&  \Big[\frac{\hbar^2 \alpha^2_{n',p}}{2MR^2}-\mu_{\sigma}\Big]\delta_{n'n''}+ \chi_0 \int r dr R_{n',p}I(r)R_{n'',p},  \\
\Omega^{n'',q}_{n',p} &=& \Omega_0 \int r dr R_{n',p}(r) I(r)R_{n'',q}(r), \\
\Delta^{n'',q}_{n',p} &=&\int r dr R_{n',p}(r)\Delta(r)R_{n'',q}(r).
\end{eqnarray}
The order parameter along the radial direction is therefore
\begin{align}
\Delta(r)  =  \frac{g}{2}\frac{1}{2\pi}\sum_{m n n'n''}\Big[
c^{(n')}_{\downarrow mn}R_{n',m+l}d^{(n'')}_{\uparrow mn}R_{n'',m+l-\kappa}\vartheta(-\epsilon_{mn})  +c^{(n')}_{\uparrow mn}R_{n',m-l}d^{(n'')}_{\downarrow mn}R_{n'',m-l-\kappa}\vartheta(\epsilon_{mn})\Big].
\end{align}
\end{widetext}

\begin{figure}[b]
\begin{center}
\includegraphics[width=0.5\textwidth]{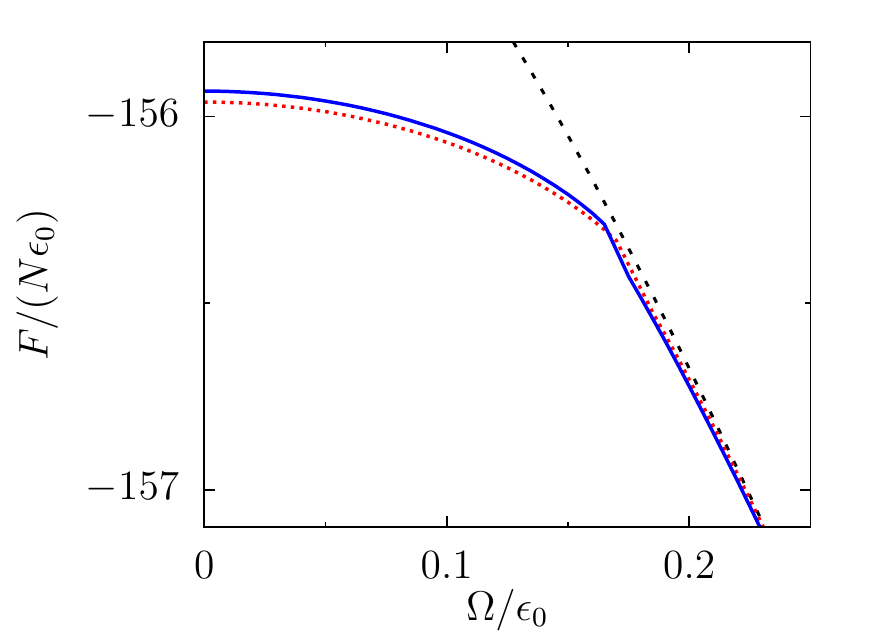}
\caption{Free energies for states with different $\kappa$. The blue solid (red dotted) curve denotes $\kappa=0$ ($\kappa=1$), and the black dashed line represents the normal state with $\Delta=0$. Here, $h/\epsilon_0=0.8$. Other parameters are the same as those in Fig.~\ref{Fig2} of the main text.}
\label{FigS1}
\end{center}
\end{figure}

\section{Free energy}
\label{app:free}
The free  energy can be obtained straightforwardly from the standard Bardeen-Cooper-Schrieffer pairing theory, with
\begin{align}
F  = & \frac{1}{2}\sum_{mn  }\epsilon_{mn}\Big[\vartheta(-\epsilon_{mn})- \sum_{\sigma}\int d {\bf r} |v_{\sigma m n}({\bf r})|^2 \Big] \nonumber\\
&+\int d{\bf r}\frac{|\Delta({\bf r})|^2}{g}+\mu N,\label{En}
\end{align}
When $\Delta({\bf r})$ and $\mu$ are self-consistently determined, the free energy can be evaluated straightforwardly.

Figure~\ref{FigS1} shows the comparison between free energies of pairing states with different $\kappa$.
For a finite two-photon detuning $h$, we can clearly see that the ground-state changes from $\kappa=0$ to $\kappa=1$ when $\Omega_0$ becomes large enough,  which means the system enters a vortex state.

\bibliographystyle{apsrev4-1}

\begin{thebibliography}{99}
\bibitem{Lin-11} Y.-J. Lin, K. Jim\'{e}nez-Garc\'{i}a, and I. B. Spielman,
Spin–orbit-coupled Bose–Einstein condensates,
Nature (London) {\bf 471}, 83 (2011).

\bibitem{Zhang-12}
P. Wang, Z.-Q. Yu, Z. Fu, J. Miao, L. Huang, S. Chai, H. Zhai, and J. Zhang,
Spin-orbit coupled degenerate Fermi gases,
Phys. Rev. Lett. {\bf 109}, 095301 (2012).

\bibitem{Zwierlein-12}
L. W. Cheuk, A. T. Sommer, Z. Hadzibabic, T. Yefsah, W. S. Bakr, and M. W. Zwierlein,
Spin-injection spectroscopy of a spin-orbit coupled Fermi gas,
Phys. Rev. Lett. {\bf 109}, 095302 (2012).

\bibitem{socreview1} V. Galitski and I. B. Spielman, 
Spin–orbit coupling in quantum gases,
Nature (London) {\bf 494}, 49 (2013).

\bibitem{socreview2} N. Goldman, G. Juzeli\={u}nas, and P. \"Ohberg, and I. B. Spielman, 
Light-induced gauge fields for ultracold atoms,
Rep. Prog. Phys. {\bf 77}, 126401 (2014).

\bibitem{socreview3} H. Zhai, 
Degenerate quantum gases with spin–orbit coupling: a review,
Rep. Prog. Phys. {\bf78}, 026001 (2015).

\bibitem{socreview4} W. Yi, W. Zhang, and X. Cui, 
Pairing superfluidity in spin-orbit coupled ultracold Fermi gases,
Sci. China Phys. Mech. Astron. {\bf 58}, 1 (2015).

\bibitem{socreview5} J. Zhang, H. Hu, X. J. Liu, and H. Pu, 
Fermi gases with synthetic spin–orbit coupling,
Annu. Rev. Cold At. Mol. {\bf 2}, 81 (2014).

\bibitem{socreview6} L. Zhang and X. J. Liu, 
Spin-orbit coupling and topological phases for ultracold atoms,
in \textit{Synthetic Spin-Orbit Coupling in Cold Atoms}, 
edited by W. Zhang, W. Yi, and C. A.R. S\'a  Melo (World Scientific, Singapore, 2018), pp.1--87.

\bibitem{puhantwobody} L. Dong, L. Jiang, H. Hu, and H. Pu, 
Finite-momentum dimer bound state in a spin-orbit-coupled Fermi gas,
Phys. Rev. A {\bf 87}, 043616 (2013).

\bibitem{shenoy} V. B. Shenoy, 
Flow-enhanced pairing and other unusual effects in Fermi gases in synthetic gauge fields,
Phys. Rev. A {\bf 88}, 033609 (2013).

\bibitem{soc3body1} Z.-Y. Shi, X. Cui, and H. Zhai, 
Universal trimers induced by spin-orbit coupling in ultracold Fermi gases,
Phys. Rev. Lett. {\bf 112}, 013201 (2014).

\bibitem{soc3body2} X. Cui and W. Yi, 
Universal Borromean binding in spin-orbit-coupled ultracold Fermi gases,
Phys. Rev. X {\bf 4}, 031026 (2014).

\bibitem{Wu-13} F. Wu, G.-C. Guo, W. Zhang, and W. Yi, 
Unconventional superfluid in a two-dimensional Fermi gas with anisotropic spin-orbit coupling and Zeeman fields,
Phys. Rev. Lett. {\bf 110}, 110401 (2013).

\bibitem{chuanwei-13}
C. Qu, Z. Zheng, M. Gong, Y. Xu, L. Mao, X. Zou, G. Guo, and C. Zhang,
Topological superfluids with finite-momentum pairing and Majorana fermions,
Nat. Commun. {\bf 4}, 2710 (2013).

\bibitem{Wei-13}
 W. Zhang and W. Yi,
 Topological Fulde–Ferrell–Larkin–Ovchinnikov states in spin–orbit-coupled Fermi gases,
 Nat. Commun. {\bf 4}, 2711 (2013).

\bibitem{tfflo0} C. Chen, 
Inhomogeneous topological superfluidity in one-dimensional spin-orbit-coupled Fermi gases,
Phys. Rev. Lett. {\bf 111}, 235302 (2013).

\bibitem{tfflohu} X.-J. Liu and H. Hu,
Topological Fulde-Ferrell superfluid in spin-orbit-coupled atomic Fermi gases,
Phys. Rev. A {\bf 88}, 023622 (2013).

\bibitem{xjliu1D} X.-J. Liu, Z.-X. Liu, and M. Cheng, 
Manipulating topological edge spins in a one-dimensional optical lattice,
Phys. Rev. Lett. {\bf 110}, 076401 (2013).

\bibitem{xjliu2D} X.-J. Liu, K.T. Law, and T.K. Ng, 
Realization of 2D spin-orbit interaction and exotic topological orders in cold atoms,
Phys. Rev. Lett. {\bf 112}, 086401 (2014).

\bibitem{Qi-16}
 L. Huang, Z. Meng, P. Wang, P. Peng, S.-L. Zhang, L. Chen,
D. Li, Q. Zhou, and J. Zhang, 
Experimental realization of two-dimensional synthetic spin–orbit coupling in ultracold Fermi gases,
Nat. Phys. {\bf 12}, 540 (2016).

\bibitem{Shuai-16}
Z. Wu, L. Zhang, W. Sun, X.-T. Xu, B.-Z. Wang, S.-C. Ji, Y. Deng, S. Chen, X.-J. Liu, and J.-W. Pan, 
Realization of two-dimensional spin-orbit coupling for Bose-Einstein condensates,
Science {\bf 354}, 83 (2016).

\bibitem{Zhang-16}
Z. Meng, L. Huang, P. Peng, D. Li, L. Chen, Y. Xu, C. Zhang, P. Wang, and J. Zhang, 
Experimental observation of a topological band gap opening in ultracold Fermi gases with two-dimensional spin-orbit coupling,
Phys. Rev. Lett. {\bf 117}, 235304 (2016).

\bibitem{Kane-08}
L. Fu and C.L. Kane, 
Superconducting proximity effect and Majorana fermions at the surface of a topological insulator,
Phys. Rev. Lett. {\bf 100}, 096407 (2008).

\bibitem{tsfsolid1} J. D. Sau, R. M. Lutchyn, S. Tewari, and S. Das Sarma, 
Generic new platform for topological quantum computation using semiconductor heterostructures,
Phys. Rev. Lett. {\bf 104}, 040502 (2010).

\bibitem{tsfsolid2} J. Alicea, 
Majorana fermions in a tunable semiconductor device,
Phys. Rev. B {\bf 81}. 125318 (2010).

\bibitem{tsfsolid3} Y. Oreg, G. Refael, and F. von Oppen, 
Helical liquids and Majorana bound states in quantum wires,
Phys. Rev. Lett. {\bf 105}, 177002 (2010).

\bibitem{tsfsolid4} L. Mao, J. Shi, Q. Niu, and C. Zhang, 
Superconducting phase with a chiral 
$f$-wave pairing symmetry and Majorana fermions induced in a hole-doped semiconductor,
Phys. Rev. Lett. {\bf 106}, 157003 (2011).

\bibitem{soc1} S. Tewari, T. D. Stanescu, J. D. Sau, and S. Das Sarma, 
Topologically non-trivial superconductivity in spin–orbit-coupled systems: bulk phases and quantum phase transitions,
New J. Phys.  {\bf 13}, 065004 (2011).

\bibitem{Yi-11}
J. Zhou, W. Zhang, and W. Yi, 
Topological superfluid in a trapped two-dimensional polarized Fermi gas with spin-orbit coupling,
Phys. Rev. A {\bf 84}, 063603 (2011).

\bibitem{Jiang-19}
D. Zhang, T. Gao, P. Zou, L. Kong, R. Li, X. Shen, X.-L. Chen, S.-G. Peng, M. Zhan, H. Pu, and K. Jiang,
Ground-state phase diagram of a spin-orbital-angular-momentum coupled Bose-Einstein condensate,
Phys. Rev. Lett. {\bf 122}, 110402 (2019).

\bibitem{Lin-18}
H.-R. Chen, K.-Y. Lin, P.-K. Chen, N.-C. Chiu, J.-B. Wang, C.-A. Chen, P.-P. Huang, S.-K. Yip, Y. Kawaguchi, and Y.-J. Lin,
Spin–orbital-angular-momentum coupled Bose-Einstein condensates,
Phys. Rev. Lett. {\bf 121}, 113204 (2018).

\bibitem{Hu-15}
Y.-X. Hu, C. Miniatura, and B. Gr\'{e}maud,
Half-skyrmion and vortex-antivortex pairs in spinor condensates,
Phys. Rev. A {\bf 92}, 033615 (2015).

\bibitem{Pu-15}
M. DeMarco and H. Pu,
Angular spin-orbit coupling in cold atoms,
Phys. Rev. A  {\bf 91}, 033630 (2015).

\bibitem{Sun-15}
K. Sun, C. Qu, and C. Zhang,
Spin–orbital-angular-momentum coupling in Bose-Einstein condensates,
Phys. Rev. A {\bf 91}, 063627 (2015).

\bibitem{Qu-15}
C. Qu, K. Sun, and C. Zhang,
Quantum phases of Bose-Einstein condensates with synthetic spin–orbital-angular-momentum coupling,
Phys. Rev. A {\bf 91}, 053630 (2015).

\bibitem{Chen-16}
 L. Chen, H. Pu, and Y. Zhang,
 Spin-orbit angular momentum coupling in a spin-1 Bose-Einstein condensate,
 Phys. Rev. A {\bf 93}, 013629 (2016).

\bibitem{Hu-19}
 X.-L. Chen, S.-G. Peng, P. Zou, X.-J. Liu, and H. Hu, 
 Angular stripe phase in spin-orbital-angular-momentum coupled Bose condensates,
 Phys. Rev. Research {\bf 2}, 033152 (2020).

\bibitem{Chen-19}
K.-J. Chen, F. Wu, J. Hu, and L. He, 
Ground-state phase diagram and excitation spectrum of a Bose-Einstein condensate with spin-orbital-angular-momentum coupling,
Phys. Rev. A {\bf 102}, 013316 (2020).

\bibitem{Han-20}
L. Chen, Y. Zhang, and H. Pu, 
Spin-nematic vortex states in cold atoms,
Phys. Rev. Lett. {\bf 125}, 195303 (2020).

\bibitem{Duan-20}
Y. Duan, Y. M. Bidasyuk, and A. Surzhykov, 
Symmetry breaking and phase transitions in Bose-Einstein condensates with spin–orbital-angular-momentum coupling,
Phys. Rev. A {\bf 102}, 063328 (2020).

\bibitem{Chen-20}
K.-J. Chen, F. Wu, S.-G. Peng, W. Yi, and L. He, 
Generating giant vortex in a fermi superfluid via spin-orbital-angular-momentum coupling,
Phys. Rev. Lett. {\bf 125}, 260407 (2020).

\bibitem{Wang-21}
 L.-L. Wang, A.-C. Ji, Q. Sun, and J. Li,
 Exotic vortex states with discrete rotational symmetry in atomic fermi gases with spin-orbital–angular-momentum coupling,
 Phys. Rev. Lett. {\bf 126}, 193401 (2021).

\bibitem{yisoc} W. Yi and G.-C. Guo, 
Phase separation in a polarized Fermi gas with spin-orbit coupling,
Phys. Rev. A {\bf 84}, 031608(R) (2011).

\bibitem{Zak-89}
J. Zak, 
Berry's phase for energy bands in solids,
Phys. Rev. Lett. {\bf 62}, 2747 (1989).

\bibitem{Wei-12}
R. Wei and E. J. Mueller, 
Majorana fermions in one-dimensional spin-orbit-coupled Fermi gases,
Phys. Rev. A {\bf 86}, 063604 (2012).

\bibitem{Liu-12}
X.-J. Liu and H. Hu, 
Topological superfluid in one-dimensional spin-orbit-coupled atomic Fermi gases,
Phys. Rev. A {\bf 85}, 033622 (2012).


\end{thebibliography}

\end{document}